\documentclass[12pt]{article}
\usepackage{epsfig}

\def\beq{\begin{equation}}
\def\eeq#1{\label{#1}\end{equation}}
\def\eeqn{\end{equation}}

\def\beqa{\begin{eqnarray}}
\def\eeqa#1{\label{#1}\end{eqnarray}}
\def\eeqan{\end{eqnarray}}

\let\bar=\overbar

\def\Dslash{\not{\hbox{\kern-4pt $D$}}}
\def\dslash{\not{\hbox{\kern-2pt $\del$}}}

\def\msb{{\bar{\ssstyle M \kern -1pt S}}}

\def\Title#1{\begin{center} {\Large {\bf #1} } \end{center}}

\def\tbr{\tau_{\rm br}}
\def\fbr{f_{\rm br}}
\def\Epk{E_{\rm pk}}
\def\Eej{E_{\rm ej}}

\def\tobs{t_{\rm obs}}
\def\GCM{{\bar\Gamma}}
\def\sT{\sigma_{\rm T}}
\def\Uf{U_{\rm free}}
\def\Erot{E_{\rm rot}}
\def\mdec{m_{\rm dec}}
\def\Rdec{R_{\rm dec}}
\def\Rgap{R_{\rm gap}}
\def\Racc{R_{\rm acc}}
\def\Rload{R_{\rm load}}
\def\Rann{R_\pm}

\def\tpeak{t_{\rm peak}}
\def\Gsh{\Gamma_{\rm sh}}
\def\Gej{\Gamma_{\rm ej}}
\def\Tann{T_\pm}
\def\dd{\rm d}

\newbox\grsign \setbox\grsign=\hbox{$>$} \newdimen\grdimen \grdimen=\ht\grsign
\newbox\simlessbox \newbox\simgreatbox \newbox\simpropbox
\setbox\simgreatbox=\hbox{\raise.5ex\hbox{$>$}\llap
     {\lower.5ex\hbox{$\sim$}}}\ht1=\grdimen\dp1=0pt
\setbox\simlessbox=\hbox{\raise.5ex\hbox{$<$}\llap
     {\lower.5ex\hbox{$\sim$}}}\ht2=\grdimen\dp2=0pt
\setbox\simpropbox=\hbox{\raise.5ex\hbox{$\propto$}\llap
     {\lower.5ex\hbox{$\sim$}}}\ht2=\grdimen\dp2=0pt
\def\simgt{\mathrel{\copy\simgreatbox}}
\def\simlt{\mathrel{\copy\simlessbox}}

\begin{document}

\Title{Prompt Emission and Early Afterglows of Gamma-Ray Bursts}

\bigskip\bigskip


\begin{raggedright}  

{\it Andrei M. Beloborodov\index{Beloborodov, A.M.}\\
Canadian Institute for Theoretical Astrophysics\\
University of Toronto, 60 St. George Street \\
Toronto, M5S 3H8 Ontario, CANADA}
\bigskip\bigskip
\end{raggedright}

\section{Introduction}

Gamma-ray bursts (GRBs) are extremely bright ($10^{50}-10^{53}$~erg/s) and 
short ($10^{-2}-10^{2}$~s) emission events observed from distant parts of 
the Universe. Their redshifts are now measured in about 20 cases and typical 
$z\sim 1$ (up to 4.5) are found~\cite{Djorg}. The burst were observed to 
occur with a rate of $\approx 1$ per day by the BATSE experiment~\cite{Fishman} 
and even a higher rate would probably be detected with more sensitive 
instruments~\cite{Stern1}. GRBs are very different from supernovae not only 
because of their short duration and high luminosity. Their most special feature 
is that the emission peaks in the gamma-ray band, at $h\nu$ of a few hundred 
keV or perhaps more. In many cases ($\approx 50$\%) the gamma-ray bursts are 
followed by afterglows --- a much longer and softer emission that deacays in 
time and evolves from X-rays (hours) to radio (months)~\cite{vanPar}.

What triggers the bursts is still uncertain. The primary 
energy release occurs in a very compact region (probably within $10^{7}$~cm 
--- from variability arguments) which is comparable to the size of 
compact objects --- black holes and neutron stars. The energy output
(assuming isotropic explosion) approaches a stellar rest-mass energy. 
This suggests a high efficiency of mass conversion into radiation and points 
to a relativistic collapse with a gravitational potential $\phi\sim c^2$ which
agrees with the potential of compact objects. Yet the configuration
of the progenitor system and the reason of the collapse are uncertain
(see reviews \cite{Piran,Mesz1} and refs. therein). 
It can be (I) coalescence of a close binary consisting of two 
compact objects~\cite{Pacz1}, (II) collapse of a massive star 
core~\cite{Woosley}, and (III) collapse of a white dwarf~\cite{Usov92}. 
Also conversion of neutron stars into strange stars was 
proposed~\cite{Cheng,Bombaci}.
Different scenarios can be tested against observations. In particular,
massive stars spend their short lifes close to where they were born,
and hence in the second scenario GRBs should occur in regions of active star 
formation. There is a growing evidence that this is indeed the 
case~\cite{Djorg}.

\begin{figure}[htb]
\begin{center}
\epsfig{file=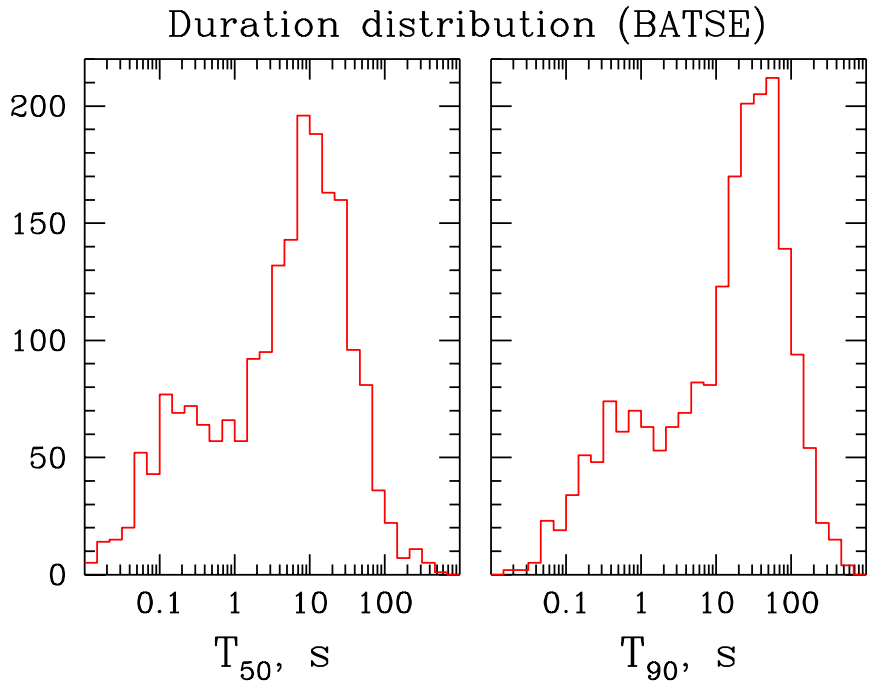,height=9cm}
\caption{The duration distribution of 2041 GRBs from the current BATSE
catalogue. Right panel: the duration is measured by T90, which is the time
over which a burst emits from 5\% of its total counts to 95\%. Left panel:
the duration is measured by $T_{50}$ --- the time
over which a burst emits from 25\% of its total counts to 75\%.
The counts are summed over all 4 BATSE energy channels (photon energy 
$E > 20$keV).}
\label{fig:dur}
\end{center}
\end{figure}

The afterglow is believed to come from a relativistic blast wave driven by 
the central explosion into an ambient medium. If GRBs are triggered by old 
stellar remnants they are likely surrounded by an interstellar medium (ISM) 
of density $n\sim 1$~cm$^{-3}$ or even lower if the remnants are kicked out 
of the disk of the parent galaxy.
In contrast, a GRB with a massive progenitor should occur inside a dense and 
strongly inhomogeneous star-forming cloud. Even more importantly, a massive 
progenitor emits a powerful wind all the time before it explodes, and this 
wind is the actual ambient medium of the GRB, with density scaling with 
radius as $R^{-2}$ out to a distance of a few parsecs~\cite{Li}. 
The afterglow observations were supposed to reveal the nature of the 
ambient medium and thereby the progenitor.
Yet after having observed many afterglows no definite conclusions are made.
Perhaps the most surprising finding is that the afterglows are very diverse.


\section{Prompt Emission: Observations}

The prompt $\gamma$-ray emission (called prompt as opposed to the afterglow) 
displays a rich phenomenology which is poorly understood from a theoretical
standpoint. In this section basic observed properties of the prompt GRBs are 
briefly summarized.

\subsection{Time profiles: preferred timescales}

The duration of GRBs varies by 4 orders of magnitude.
The duration distribution is shown in Fig.~1. 
It has three clear features:
(1) a break at short $T_{50}\approx 0.1$~s, 
(2) a break at long $T_{50}\approx 10$~s, and 
(3) a deficit of GRBs with $T_{50}\approx 1$~s. 
The three timescales are not explained yet.

Time profiles $C(t)$ of GRBs are very diverse. They show strong variations on all 
resolved timescales (the resolution $\Delta t=64$~ms in the standard 
BATSE data). In long bursts the range $\Delta t<t<T_{50}$ spans three decades
and Fourier transform $C(f)$ of the profile is useful: its power spectrum 
$P_f=C_fC_f^*$ shows how the variability power is distributed over timescales.  
It reveals an interesting special feature of GRBs:
$P_f$ follows a power law with slope $\alpha\approx- 5/3$ at frequencies 
$(2\pi T_{50})^{-1}<f<\fbr\approx 1$~Hz and breaks at $\fbr$ (Fig.~2). 
This Fourier spectrum is observed in individual GRBs and it becomes especially 
clear when $P_f$ of several GRBs are averaged: then the statistical fluctuations 
$\Delta P_f/P_f\sim 1$ are averaged out and a perfect power law with a sharp 
break is found~\cite{Bel1}.
The break is not an artifact of the 64~ms time binning; it rather indicates a 
timescale $\tbr\approx (1/2\pi)$~s below which the variability decreases. 

Various techniques were used to study the GRB profiles, for example
decomposition into separate pulses~\cite{Norris}, 
construction of the average time profile~\cite{Stern1}, and 
construction of the average auto-correlation function (ACF)~\cite{Fenimore}. 
The results of different approaches are related. For example, the pulse 
decomposition gives a distribution of pulse widths with a break at a 
fraction of second, in agreement with the break in the power spectrum $P_f$. 
The ACF is just a Fourier transform of $P_f$ and it is perfectly 
fitted as a stretched exponential $\exp[-(\tau/\tau_0)^\beta]$ of index 
$\beta=\alpha-1$~\cite{Bel1}. The ACF width $\tau_0$ is about the geometrical 
mean of $\tbr$ and $T_{50}$.


\begin{figure}[t]
\begin{center}
\epsfig{file=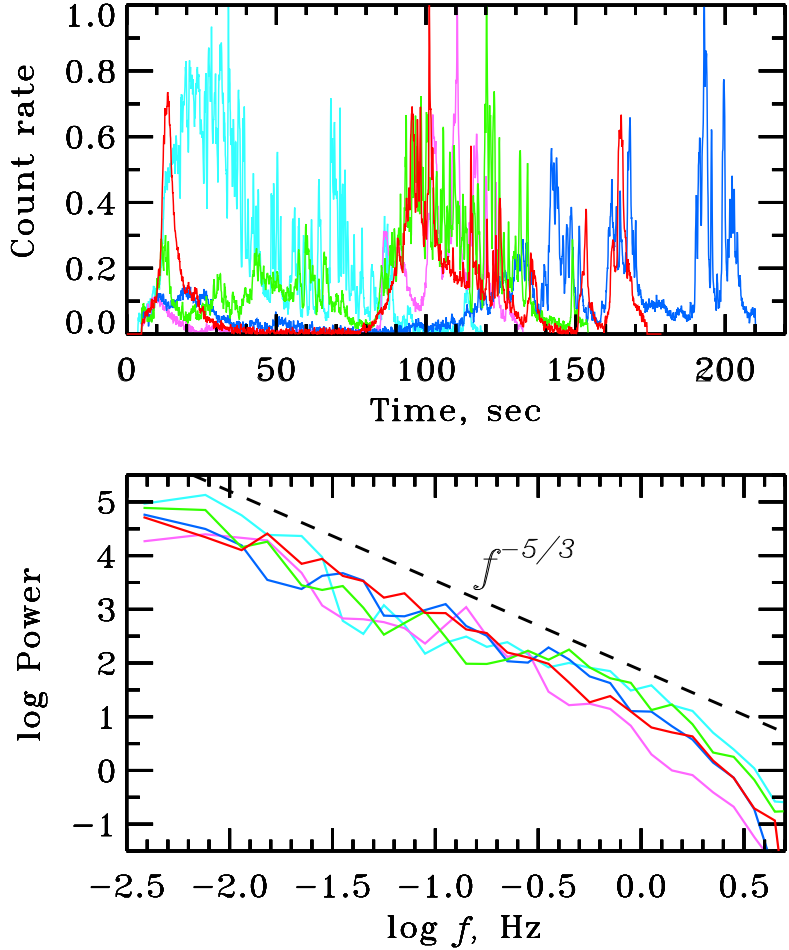,height=12.7cm}
\caption{Peak-normalized time profiles and their 
power spectra for the 5 brightest BATSE bursts with $T_{90}>100$~s 
(brightness is measured by the peak count rate).}
\end{center}
\end{figure}

\subsection{Spectra: a preferred photon energy?}

The GRB energy spectrum is very difficult to study because it varies rapidly 
during the burst. The time-averaged spectrum is well fitted by a (smoothly)
broken power law with a low-energy slope $\Gamma_s=-1\pm0.5$ and a high-energy 
slope $\Gamma_h=-2.3\pm 0.5$~\cite{Preece}. The spectral peak position $\Epk$ 
was found to cluster around $\sim 200$~keV. This clustering is not understood.
Theoretically, there is no convincing model that would reproduce it.
Observationally, one may ask whether the results based on the BATSE data are 
conclusive, since the experiment had the highest sensitivity at $\sim 100$~keV. 
Statistical studies point toward reality of the $\Epk$ 
clustering~\cite{Preece}, yet a confirmation by a future mission is desirable.
One should also keep in mind that the time-average spectrum is normally
far from the instantaneous one and its parameters do not
give a direct information on the emission mechanism. 
Indeed, the instantaneous spectrum was fitted by a similar function yet with
different (evolving) $\Epk$ and slopes. $\Epk$ can change during one burst 
by more than one order of magnitude. An interesting feature
of spectral evolution is a positive correlation between the 
instantaneous $\Epk$ and the energy flux $F$. In particular, the dependence
$F\propto\Epk^\gamma$ was found for the decay phase of GRB 
where $\gamma=1-3$ varies from pulse to pulse and the average 
$\bar\gamma\approx 2$~\cite{Borgonovo}.
 
The question of preferential energy can be reformulated:
do GRB-like bursts occur in different energy bands? Recently, such bursts 
(X-ray flashes) were found by BeppoSax~\cite{Heise}. These events have 
durations and time profiles similar to classical GRBs, but they emit most of 
their energy at softer energies (below 20~keV). The hardness ratio distribution 
of these soft bursts indicates that they extend continuously the population of 
the classical GRBs to lower energies.


\section{Models for Prompt Emission}

A basic constraint on GRB models is derived from their fast variability
and enormous power: GRBs must be emitted by highly relativistic ejecta 
directed toward the observer. A lower bound on the ejecta Lorentz factor 
$\Gamma$ can be estimated as follows. Baryonic ejecta with kinetic luminosity 
$L_k\sim 10^{52}$~erg/s become transparent to electron scattering at a radius 
$R_*=L_k\sT/(8\pi m_p c^3\Gamma^3)\sim 10^{19}\Gamma^{-3}$~cm.
The observed emission from the ejecta can vary on timescales 
$t_v\geq R_*/(2\Gamma^{2}c)\sim 10^8\Gamma^{-5}$~s. 
The actual $t_v<0.1$~s then requires $\Gamma\sim 10^2$. 

A simple phenomenological model of a relativistic outflow envisions a
central engine of size $r_0\sim 10^7$~cm which deposits energy at a high rate
$\sim 10^{52}$~erg/s at the base of the outflow; this causes high pressure
and free expansion with increasing $\Gamma\propto R$~\cite{Pacz2}.
The engine operates much longer ($\tobs\sim$~s) than its light-crossing time 
($r_0/c\simlt$~ms).  It ejects a long sequence of blobs $\sim r_0$ that undergo
free expansion (on timescale $r_0/c$) one after another and form a continuous 
outflow.

The high $\Gamma$ of the outflow implies that energy is ejected with 
a very low baryon ``pollution''. This energy must be carried by a highly
relativistic plasma and frozen magnetic fields. 
Below we first describe outflows dominated by the plasma energy 
and then those dominated by magnetic energy.

\subsection{Baryonic Outflow: Internal Shocks}

Suppose the engine emits thermal energy into a solid angle $\Omega$ with rate 
$L(\Omega/4\pi)$. The energy density at the outflow base, $w=L/(4\pi r_0^2c)$,
is so high that the matter must be in complete thermodynamic equilibrium with 
radiation at a temperature $kT_0\sim$~MeV. Here the matter is strongly 
dominated by $e^\pm$ pairs which maintain an equilibrium with radiation
through reaction $\gamma+\gamma\leftrightarrow e^++e^-$, and $w$ is 
dominated by photons and $e^\pm$~\cite{Cavallo}. 
The high-pressure material expands freely with acceleration. The expansion 
is accompanied by adiabatic cooling, eventually the temperature drops, and 
pairs annihilate. 

A crucial parameter of the outflow is its baryonic rest-mass luminosity, $L_b$.
If there are no baryons ($L_b=0$) the outflow becomes transparent after 
$e^\pm$ annihilation and the radiation escapes. This happens when the
temperature in the comoving frame drops to $k\Tann\sim 10$~keV, at a radius 
$R=r_0(T_0/\Tann)$ 
and $\Gamma=(T_0/\Tann)$. Essentially all the explosion energy is carried away 
by the blackbody radiation with temperature $T=\Gamma \Tann=T_0$. Thus,
a strong burst is produced at MeV energies.
Only a small energy remains available  for an afterglow; it is carried by 
an optically thin outflow of surviving $e^\pm$ with luminosity
$L_\pm=8\pi (r_0m_ec^3/\sT)(T_0/\Tann)^4\ll L$.
In contrast, the observed afterglows
are comparable in energy with the prompt GRB, and this is one reason why the 
outlined model is unfavorable. Besides, the blackbody exponential cutoff 
was not found in GRB spectra and a nonthermal emission mechanism is preferred.

A small baryonic pollution, $L_b\ll L$, can drastically change the model
in that it keeps the outflow optically thick after $e^\pm$ annihilation. 
The radiation remains trapped, continues to accelerate 
the outflow, $\Gamma\propto R$, and cools down adiabatically.
The kinetic luminosity of baryons $L_k=\Gamma L_b$ grows and approaches $L$
at $R_s=(L/L_b)r_0$. Now the explosion luminosity has been converted into
$L_k$ and the Lorentz factor saturates at $\Gamma=L/L_b$.

The baryonic outflow was suggested to emit GRBs by internal 
shocks~\cite{Rees,Kob1,Daigne}. If $L_b/L$ varies during 
the outflow ejection, $\Gamma$ will fluctuate, which results in internal 
collisions (caustics). A minimum scale of the $\Gamma$-fluctuations is 
$\lambda_0\sim r_0$ --- the size of the central engine. 
First internal collisions will occur at a radius $R_0=2\GCM^2\lambda_0/A$ 
where $A=\Delta\Gamma/\GCM$ is the fluctuation amplitude (rms) and $A<1$ is 
assumed~\cite{Bel2}.
At $R>R_0$ the collisions proceed on scales $\lambda>\lambda_0$ in 
a hierarchical manner. For a given initial spectrum of fluctuations one can 
compute the history of internal dissipation~\cite{Bel2,Bel3}.

The transparency radius of the baryonic outflow $R_*=L\sT/(8\pi m_pc^3\Gamma^3)$ 
is comparable to $R_0$, and hence internal shocks can emit nonthermal spectra. 
Any time profiles of GRBs can be fitted by the model with properly chosen 
initial conditions. One would also like to know what fraction $\epsilon$ 
of the total bulk kinetic energy $E$ is dissipated internally.
It is determined by an initial free energy $\Uf$ of the fluctuating outflow:
$\epsilon=\Uf/E$. $\Uf$ is easily calculated analytically given an initial 
amplitude of fluctuations $A<1$, leading to $\epsilon=A^2/2<0.5$~\cite{Bel2}.
The case of $A>1$ is also possible and it is more complicated. At high $A$
the Lorentz factors of colliding shells differ by a big factor 
$\Gamma_2/\Gamma_1$ up to $10^2-10^3$; then the outflow evolves in a non-linear
way and $\epsilon$ could approach unity. A realistic 
$\Gamma_2/\Gamma_1$ is limited because a radiation-pressure-accelerated 
ejecta has $\Gamma\leq(R/r_0)$ at any given $R$. A slow shell with $\Gamma_1$ 
is hit by a following fast shell with $\Gamma_2\gg\Gamma_1$ at 
$R_0=2\Gamma_1^2\lambda_0$ (where $\lambda_0$ is their initial separation);
since $\Gamma_2\leq R_0/r_0$ one gets a maximum 
$\Gamma_2/\Gamma_1=2\Gamma_1(\lambda_0/r_0)$. It may be much higher 
than $2\Gamma_1$ if the central engine ``waits'' before ejecting a very
energetic shell ($\lambda_0\gg r_0$) and the shell has space ahead to
develop a high $\Gamma_2$ before colliding. Therefore, in the non-linear regime,
$\epsilon$ is not a simple function of $A$ and it is sensitive to details of 
initial conditions. Additional complications are owing to possible 
$e^\pm$ production~\cite{Guetta} which depends on the unknown spectrum of the 
produced $\gamma$-rays at $h\nu>100$~MeV.

A major uncertainty of this scenario is the emission mechanism of internal shocks
which remains to be a matter of controversy. The simplest model of synchrotron
emission from shock-accelerated electrons predicts a broken power-law radiation 
spectrum with a low energy slope $\Gamma_s>-2/3$~\cite{Piran} and this is 
inconsistent with the data~\cite{Preece}. Possible modifications were 
discussed~\cite{Ghisellini,Mesz2,Lloyd} yet the issue is not yet settled. 
Also the radiative efficiency $\eta$ of an internal shock is unknown and
speculations range from $\eta\ll 1$ to $\eta=1$. The net radiative efficiency 
of the outflow can be high even when $\eta\ll 1$ in each shock~\cite{Kob2}.
The fraction of the emitted radiation that appears in the BATSE spectral window 
depends on the model assumptions and may be small~\cite{Spada}.


\subsection{Magnetic Outflow: Field Dissipation}


The central compact engine is likely to have a rotational energy 
$E_{\rm rot}$ comparable to the gravitational energy. Differential rotation 
can generate very strong magnetic fields, as high as $B\sim 10^{16}$~G. 
For example, neutron stars can be born strongly magnetized~\cite{Usov92,
Thomp1}. When a compact binary merges or a massive star core collapses, 
a black hole forms and accretes debris. Then strong magnetic fields can be 
generated by the differentially rotating accretion disk.
A magnetized rotator emits Poynting flux with luminosity 
$L_P\approx\mu^2\omega^4/c^3$ where $\mu$ is the magnetic moment and 
$\omega$ is the angular velocity of rotation. For expected 
$\mu\sim 10^{34}$~G~cm$^3$ and $\omega\sim 10^4$~s$^{-1}$ one gets 
$L_P\sim 10^{52}$~erg/s
and hence $E_{\rm rot}$ is emitted in a few seconds. The resulting 
Poynting-flux-dominated outflow was studied as a possible GRB source;
we will describe here the scenario proposed in~\cite{Usov94}. 

A fraction $\sigma$ of $\Erot$ can be dissipated at the base of the Poynting 
outflow, loading it with baryon-free $e^\pm$ plasma and trapped blackbody 
radiation. The Poynting flux is ``frozen'' into the plasma whose early 
dynamics is similar to the scenario described in Section~3.1. The difference 
is that now only a fraction $\sigma$ of the total energy is thermalized and 
the initial temperature $T_0$ is lower by a modest factor $\sigma^{1/4}$. 
The magnetized outflow is accelerated by the thermal pressure until the 
temperature drops to $k\Tann\sim 10$~keV. Here pairs annihilate to 
transparency, the thermal radiation decouples, and the Lorentz factor 
of the surviving $e^\pm$ plasma saturates at $\Gamma=T_0/\Tann\sim 10^2$.

The coasting $e^\pm$ outflow carries the initially frozen magnetic field as 
long as the particle density $n$ is sufficient to ensure the MHD approximation.
The toroidal field component $B_\phi$ is least suppressed in the expansion 
process and $B\approx B_\phi\propto R^{-1}$. The magnetic outflow (like the 
baryonic one) can be inhomogeneous on scales $\lambda\simgt r_0$, so that 
the field possesses a free energy $\Uf$ of currents 
$j=(c/4\pi)|\nabla\times{\bf B}|\approx (c/4\pi)(B/r_0)\propto R^{-1}$. 
The density of surviving $e^\pm$ at $R>\Rann$ is 
$n_+\sim\Gamma\Rann/(\sT R^2)\propto R^{-2}$ and 
outside $\sim 10^{14}$~cm it is too small to support the currents. Then
the displacement current $\dot{E}/c=j$ is generated and $\Uf$ is converted 
into large-amplitude electromagnetic waves. 
The waves have a low frequency, $c/r_0\sim 10^4$~Hz, and they are quickly
absorbed by the $e^\pm$ plasma, resulting in particle acceleration to a Lorentz 
factor $\gamma\sim L_P/L_\pm\sim 10^6$. The accelerated $e^\pm$ move in 
the residual magnetic field and the field of waves and emit photons of energy 
$h\nu\sim\gamma^2\hbar(c/r_0)\sim$~MeV. This mechanism may give rise to 
a GRB. Very strong bursts can naturally be produced since the Poynting outflow
carries a large energy $E_{\rm rot}$ and $\Uf\simlt E_{\rm rot}$ is possible.
Like the internal shock model, GRBs with arbitrary time profiles can be 
generated, depending on the initially frozen field structure in the outflow.
The predicted energy spectrum is uncertain. After the prompt GRB 
(dissipation of $\Uf$) the magnetic fields outflows further with energy 
$\Eej=\Erot-\Uf$ until it passes the energy to an ambient medium. Here it 
gives rise to an afterglow.


\section{Afterglow}

There are a number of observations of GRB afterglows in X-rays, optical,
IR, and radio (see~\cite{vanPar} for a review). In many cases the afterglow 
spectral flux decays as a simple power law, 
$F_\nu(t)\propto\nu^{-\alpha}t^{-\beta}$.
Clear deviations from this law are also observed, e.g. breaks or humps.
The indexes $\alpha$ and $\beta$ change from burst to burst and can also 
change during one afterglow.
Important early afterglow observations in X-rays were done by 
BeppoSax~\cite{Frontera}, SIGMA~\cite{Tkach}, and BATSE~\cite{Giblin}.
These observations showed that the afterglow starts immediately after or even 
overlaps with the prompt GRB. Chandra, BeppoSax, and ASCA detected spectral 
lines of iron in the afterglow, which places strong constraints on the GRB 
progenitors~\cite{Vietri,Mesz1}.

\subsection{Standard Afterglow Model}

The energy source of the afterglows is easily explained.
The ejecta with Lorentz factor $\Gamma$ runs into an ambient medium and its 
energy is dissipated when it sweeps a sufficient ambient mass $m$. Namely, 
half of the ejecta energy $\Eej$ is dissipated when the swept inertial mass 
$m\Gamma$ (measured in the ejecta frame) reaches that of the ejecta itself, 
$\Eej/(c^2\Gamma)$. The swept mass at a radius $R$ is given by 
$m(R)=\int_0^R4\pi R^2\rho\dd R$ where $\rho(R)$ is the medium mass density. 
The characteristic 
$\mdec=\Eej/(c^2\Gamma^2)$ corresponds to a radius $\Rdec$. For example,  
$\Rdec=(3 \Eej/4\pi\rho c^2\Gamma^2)^{1/3}$ if $\rho(R)=const$. In the  
massive progenitor scenario, $\rho(R)=\dot{M}/(4\pi R^2w)$ where $\dot{M}$
and $w$ are the mass loss and velocity of the progenitor wind. In this case
$\Rdec=\Eej w/(\Gamma^2\dot{M}c^2)$. For the most likely Wolf-Rayet progenitors 
with $\dot{M}\sim 10^{-5}M_\odot$~yr$^{-1}$ and $w\approx 10^8$~cm/s, one 
gets $\Rdec\approx 10^{15}(\Eej/10^{53})$~cm. 
(As we discuss in Section~4.2, this standard estimate neglects the impact 
of the $\gamma$-ray front and turns out invalid.)

The ejecta dynamics at $R>\Rdec$ depends on what happens faster ---
the dissipated heat is radiated away or the ejecta expands by a factor of two. 
If radiative losses dominate, $\Eej$ is converted into radiation at $R\sim\Rdec$ 
and $\Gamma$ quickly decreases: $\Gamma\approx\Gej(m/\mdec)^{-1}$ at 
$m>\mdec$~\cite{Blandford}.
If expansion is faster, the heat converts back into the bulk kinetic energy via 
adiabatic cooling; then $\Gamma\approx\Gej(m/\mdec)^{-1/2}$.

The decelerating shell is made of the ejecta material and the swept ambient mass, 
with a contact discontinuity between them. The shell has a high pressure that 
drives a forward shock into the medium. (A collisionless shock is
expected to form if the Larmor radius of the reflected upstream ions 
$r_L=\Gamma^2 m_p c^2/eB_{\rm up}<R$; this condition may be not satisfied  
and then a different model is needed~\cite{Smolsky}.) 
The emission from the forward shock gives rise to the observed afterglow.
A simple emission model can fit the observations~\cite{Sari1,vanPar}. 
It assumes that a fraction $\epsilon_e$ of the shocked ion energy 
goes to the electrons and accelerates them with a power-law energy spectrum 
$\dd N/\dd E_e\propto E_e^{-p}$. The electrons emit a synchrotron spectrum 
in a magnetic field behind the shock. The field is assumed to be strongly 
amplified compared to the upstream field and it is parametrized by 
$\epsilon_B=(B^2/8\pi)w_{\rm th}^{-1}$ where $w_{\rm th}$ is the energy 
density of the shocked medium. The model has six main fitting parameters 
and they are found to vary strongly from burst to burst~\cite{Panaitescu}.
Therefore it is not obvious whether the assumptions are justified by the 
data and the afterglow physics is well captured by the model.
In any case the fits by the model provide a form for data representation;
it is used commonly and referred to as a standard afterglow model.

The breaks sometimes observed in the afterglow light curves are thought to
be indications of beaming of the GRB ejecta: a break should happen when
$\Gamma^{-1}$ equals the beaming angle $\theta_{\rm ej}$~\cite{Rhoads}. 
The beaming implies that the total GRB energy $E$ is reduced by a factor of
$\theta_{\rm ej}^2/(4\pi)$ compared to what one deduces assuming isotropy. 
Intriguingly, when this correction is applied to GRBs with evaluated 
$\theta_{\rm ej}$, $E$ displays a much smaller dispersion and clusters around 
$\sim 10^{51}$~erg~\cite{Panaitescu,Frail1}.

The deceleration radius $\Rdec$ is especially interesting since here the 
main afterglow is emitted. The afterglow should peak early, at an observed 
time $\tobs=\tpeak=\Rdec/(2\Gamma^2 c)$, and at $\tobs>\tpeak$ the emission 
gets weaker and softer because it comes from the decelerated shell. 
Unfortunately, there are no direct observations of $\Rdec$. 
In one case the angular size of the afterglow source was inferred from radio 
observations at $\tobs\approx 1$~month~\cite{Frail2}: the scintillation 
amplitude died out at that time, giving an angular size that corresponds to 
the source radius $R\approx 10^{17}$~cm, in agreement with model expectations. 
Extrapolating $R(t)$ back in time one 
finds that $\Rdec$ is likely to be within $10^{16}$~cm.
The $\tpeak$ should then be comparable to the prompt GRB duration and 
this agrees with the reported observations of hard X-ray
afterglows already at the end of the prompt GRB~\cite{Frontera,Tkach,Giblin}.
These observations revealed a spectrally distinct and smoothly decaying emission
component at the end of the burst and it was interpreted as the beginning
of an afterglow. The standard model predicted this early afterglow in hard X-rays.

\subsection{Revision of the Early Afterglow}

Recently the physics of the early afterglow was revised.
The new effect discovered is the strong impact of the prompt $\gamma$-rays 
on the afterglow blast wave. The presence of the violent radiation front 
ahead of the ejecta was neglected by the standard model while in fact it
crucially changes the medium faced by the ejecta: the leading $\gamma$-rays 
scatter off the medium, load it with $e^\pm$ pairs, and preaccelerate to a large 
Lorentz factor $\gamma$~\cite{Thomp2,Mesz3,Bel4}. These effects introduce 
three new characteristic radii into the problem: 
$\Rgap<\Racc<\Rload$~\cite{Bel4}. 
At $R<\Rgap\approx 0.3\Racc$ the $\gamma$-ray front sweeps the ambient medium out 
with $\gamma>\Gamma$ and opens a gap between the ejecta and the surfing medium. 
When the front passes radii 
$\Rgap<R<\Racc=0.7\times 10^{16}(E_\gamma/10^{53})^{1/2}$~cm 
($E_\gamma$ is the isotropic energy of the prompt GRB emission)
$\gamma$ is decreasing from $\Gamma$ to unity and the ejecta is sweeping the 
preaccelerated and $e^\pm$ loaded medium behind the radiation front. 
At $\Racc<R<\Rload\approx\sqrt{5}\Racc$ the ejecta sweeps a static 
($\gamma\approx 1$) medium dominated by loaded $e^\pm$.
Not until $R>\Rload=1.6\times 10^{16}(E_\gamma/10^{53})^{1/2}$~cm the $e^\pm$ 
loading is shut down and the standard afterglow model applies.

The afterglow should start at $R=\Rgap$ when the gap is closed and a blast 
wave is formed. Initially, the shock Lorentz factor $\Gsh\approx\Gamma/\gamma$ 
is modest and $e^\pm$ density of the ambient medium exceeds the ion density by 
a very large factor $n_+/n_i\approx 10^3$. The energy per shocked $e^\pm$ 
particle is suppressed $\propto\gamma^{-1}(n_i/n_+)$, resulting in soft emission. 
In contrast to the standard model, one expects to see the very beginning of the 
afterglow in optical/UV. As the ejecta approaches $\Racc$ the preacceleration 
$\gamma$ falls down steeply, $\Gsh$ rises, the emission hardens fast, and 
the afterglow appears in the X-ray band. This unusual soft-to-hard evolution 
of the early afterglow was detected in GRB~910402~\cite{Tkach}.
At $R>\Rload$ the front effects become small and the afterglow gradually 
fades according to the standard model. The whole afterglow 
evolution is shown in Figure~3.


\begin{figure}
\begin{center}
\epsfig{file=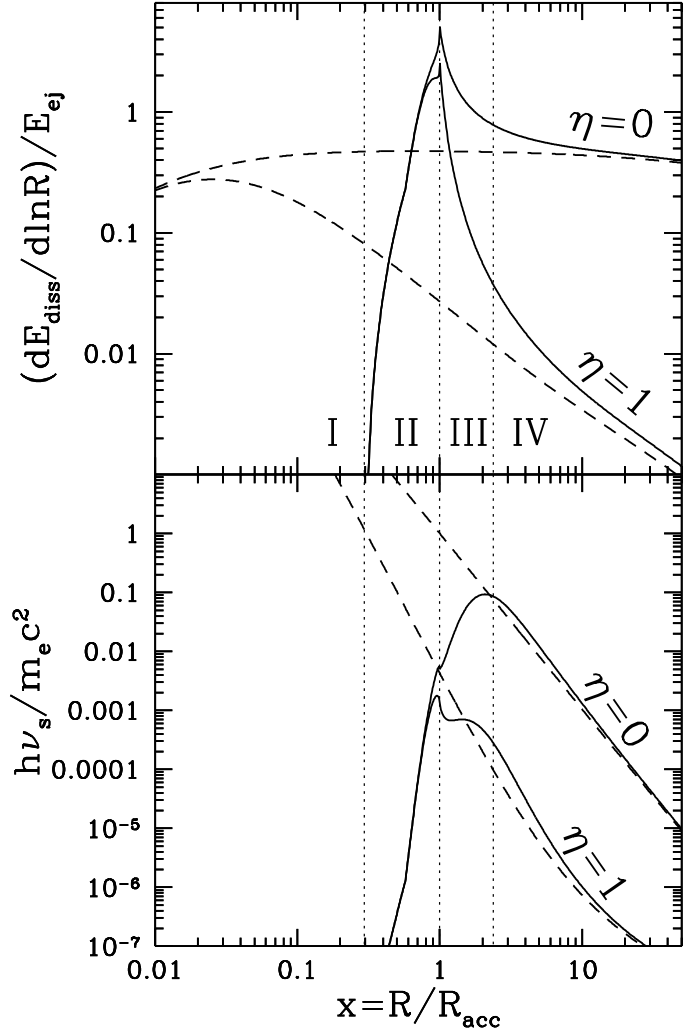,height=14cm}
\caption{ Afterglow from a GRB ejecta decelerating in a wind of a Wolf-Rayet 
progenitor with $\dot{M}=2\times 10^{-5}M_\odot$~yr$^{-1}$,
$w=10^{3}$~km~s$^{-1}$.
The GRB is modeled as an impulsive emission of a gamma-ray front
(with isotropic energy $E_\gamma=10^{53}$~erg) and a thin ejecta shell
with kinetic energy $\Eej=10^{53}$~erg and Lorentz factor $\Gej=200$.
Dashed curves show the prediction of the standard model that neglects
the impact of the radiation front and solid curves show the actual
behavior. Two extreme cases are displayed in the figure: $\eta=0$ (adiabatic
blast wave) and $\eta=1$ (radiative blast wave). Four zones are marked:
I --- $R<\Rgap$ (the gap is opened),
II --- $\Rgap<R<\Racc$ (the gap is closed and the ejecta sweeps the
relativistically preaccelerated $e^\pm$-loaded ambient medium),
III --- $\Racc<R<\Rload$ ($e^\pm$-loaded ambient medium with
$\gamma\approx 1$), and IV --- $R>\Rload$ (pair-free ambient medium
with $\gamma\approx 1$). Radius is measured in units of
$\Racc=0.7\times 10^{16}(E_\gamma/10^{53})^{1/2}$~cm.
{\it Top panel}: the dissipation rate. {\it Bottom panel}: the synchrotron
peak frequency $\nu_s$ (assuming $\epsilon_B=0.1$) in units of $m_ec^2/h$.
}
\label{fig:magnet}
\end{center}
\end{figure}

The $\gamma$-ray front can strongly impact the deceleration 
radius~\cite{Bel4}, especially in the massive progenitor scenario where the 
standard model predicts $\Rdec\ll\Racc$. The ejecta cannot decelerate 
until it approaches $\Racc$, and one finds $\Rdec\approx\Racc$.
Then the peak of the afterglow is predicted at 
$\tpeak=\Racc/(2\Gamma^2c)=12(E_\gamma/10^{53})^{1/2}(\Gamma/100)^{-2}$~s. 
A measured $\tpeak$ and $E_\gamma$ would allow one to evaluate $\Gamma$.
The isotropic energy $E_\gamma$ is now known for about 20 bursts, however
$\tpeak$ was not determined. The peak can be clearly observed in soft bands 
where the prompt GRB is not seen.
In short GRBs with duration $\ll\tpeak$ the afterglow should also be 
separated from the prompt emission in time, and the peak could be noticed even 
in the BATSE data (at $h\nu>20$~keV). A possible detection of such a peak
at tens of seconds (an excess in a time-averaged profile of 76 short GRBs) 
was recently reported~\cite{Lazzati}. 

The early soft afterglow can be studied in optical, UV, and 
soft X-ray bands on timescales less than one minute.
Normally, GRB observations in soft bands start hours or days after the 
prompt burst when only the decaying emission from $R>\Rdec$ can be observed.
Attempts were made to detect prompt optical emission at $\tobs\simgt 10$~s 
and in one case (GRB~990123) such emission was detected~\cite{Akerlof}. 
The observed optical flash was expected to come from a reverse shock in a 
baryonic ejecta~\cite{Mesz4,Sari2}. 
In~\cite{Bel4} an alternative interpretation was suggested: 
the soft flash is produced by the forward shock at its early stage when it 
propagates in the preaccelerated and $e^\pm$-loaded environment. In contrast to
the reverse shock interpretation, the ejecta is not required to be baryonic 
--- it may be a Poynting flux as well.

Prompt observations of GRBs in soft bands can
help to understand the main early phase of the ejecta-medium interaction.
A major future GRB mission is Swift (to be launched in 2003). It will provide 
optical/UV and soft X-ray data 20--70~s after the burst detection in hard X-rays.
Unfortunately, this time may be too long to catch the early soft phase of 
afterglows. Quicker observations can be done with the proposed microsatellite 
ECLAIRs~\cite{ECLAIRs} which is devoted specifically to prompt GRB observations 
in optical and soft X-rays.





\section{Concluding Remarks}

Owing to the recent observational progress, the state of the GRB field has
crucially transformed. In the beginning of 90s, $\sim 10^2$ theories did not 
contradict the data and were considered as principally possible; to a large 
extent the choice of a plausible theory was a matter of taste. The present 
situation is just opposite. GRBs have shown a mysterious complicated
phenomenology with a number of well formulated observational facts.
A successful theory is expected to make specific predictions that agree with 
the data on a quantitative level. The understanding of GRBs is far from such 
an ideal state, and the existing theories are rather naive and lacking a 
predictive power. Where models can fit the data (e.g. afterglow light curves) 
it is partially due to a large number of free parameters which ensure a 
sufficient flexibility of the model, and it does not make one confident that
the simplifying assumptions are correct.

The difficulty of the GRB theory is well compensated by the recent 
observations which put more and more constraints. The observational progress
may become even more impressive with additional channels of information such
as neutrino and gravitational radiation. New exciting observations of 
electromagnetic radiation are expected from future missions --- Swift and GLAST.
It allows one to hope for a future theory that would clarify the physics of the 
explosion and answer the basic questions:
(1) What is the progenitor and why does it collapse? (2) Where is the released 
gravitational energy channeled to before it starts to feed the outflow 
(energy of rotation, magnetic energy, heat?) 
(3) What is the composition of the ejecta ($e^\pm$, $p$, $n$, magnetic field?) 
and why are they so highly relativistic?


\def\Discussion{
\setlength{\parskip}{0.3cm}\setlength{\parindent}{0.0cm}
     \bigskip\bigskip      {\Large {\bf Discussion}} \bigskip}
\def\speaker#1{{\bf #1:}\ }
\def\endDiscussion{}

\end{document}